\documentclass{aa}

\usepackage{graphicx}
\usepackage[varg]{txfonts}
\usepackage[colorlinks=true, allcolors=blue]{hyperref}
\usepackage{amsmath}
\usepackage{amssymb}
\usepackage{textcomp}
\usepackage{gensymb}
\usepackage{mathtools}
\usepackage{subcaption}
\usepackage{lscape}
\usepackage{rotating}

\makeatletter
\renewcommand*\aa@pageof{, page \thepage{} of \pageref*{LastPage}}
\makeatother

\begin{document} 

\title{Metrewave Galactic Plane with the uGMRT (MeGaPluG) Survey: Lessons from the Pilot Study}
\titlerunning{MeGaPluG Survey}

\author{R. Dokara\inst{1}, 
        Nirupam Roy\inst{2},
        Karl Menten\inst{1}, 
        Sarita Vig\inst{3},
        Prasun Dutta\inst{4},
        Henrik Beuther\inst{5},
        Jagadheep D. Pandian\inst{3},
        Michael~Rugel\inst{6,7},
        Md~Rashid\inst{2},
        \and
        Andreas~Brunthaler\inst{1}}
\authorrunning{Dokara et al.}

\institute{Max Planck Institute for Radioastronomy (MPIfR), Auf dem H\"ugel 69, 53121 Bonn, Germany\\
           \email{rdokara@mpifr-bonn.mpg.de}
           \and
           Department of Physics, Indian Institute of Science, Bengaluru, India 560012
           \and
           Department of Earth and Space Science, Indian Institute of Space Science and Technology, Thiruvananthapuram, India 695547
           \and
           Department of Physics, Indian Insitute of Technology (Banaras Hindu University), Varanasi 221005, Uttar Pradesh, India
           \and
           Max Planck Institute for Astronomy, K\"onigstuhl 17, 69117 Heidelberg, Germany
           \and
           Center for Astrophysics, Harvard \& Smithsonian, 60 Garden St., Cambridge, MA 02138, USA
           \and
           National Radio Astronomy Observatory, 1003 Lopezville Rd, Socorro, NM 87801, USA
           }

\date{Submitted on 30 June 2023; revised 02 August 2023; accepted on 07 August 2023}

\abstract
{
The advent of wide-band receiver systems on interferometer arrays enables one to undertake high-sensitivity and high-resolution radio continuum surveys of the Galactic plane in a reasonable amount of telescope time.
However, to date, there are only a few such studies of the first quadrant of the Milky Way that have been carried out at frequencies below 1 GHz.  The Giant Metrewave Radio Telescope (GMRT) has recently upgraded its receivers with wide-band capabilities (now called the uGMRT) and provides a good opportunity to conduct high resolution surveys, while also being sensitive to the extended structures.
}
{
We wish to assess the feasibility of conducting a large-scale snapshot survey, the Metrewave Galactic Plane with the uGMRT Survey (MeGaPluG), to simultaneously map extended sources and compact objects at an angular resolution lower than $10''$ and a point source sensitivity of 0.15~mJy\,beam$^{-1}$.
}
{ 
We performed an unbiased survey of a small portion of the Galactic plane, covering the W43/W44 regions ($l=29\degree-35\degree$ and $|b|<1\degree$) in two frequency bands: 300$-$500 MHz and 550$-$750 MHz.  The 200~MHz wide-band receivers on the uGMRT are employed to observe the target field in several pointings, spending nearly 14 minutes on each pointing in two separate scans.  We developed an automated pipeline for the calibration, and a semi-automated self-calibration procedure is used to image each pointing using multi-scale CLEAN and outlier fields. 
}
{ 
We produced continuum mosaics of the surveyed region at a final common resolution of $25''$ in the two bands that have central frequencies of 400 MHz and 650 MHz, with a point source sensitivity better than 5 mJy\,beam$^{-1}$.  
A spectral index map is also obtained, which is helpful to distinguish between thermal and nonthermal emission.  Comparing with other surveys, we validated the positions and flux densities obtained from our data.  We plan to cover a larger footprint of the Galactic plane in the near future based on the lessons learnt from this study.
}
{}


\maketitle

\section{Introduction}
The Galactic mid-plane of the Milky Way is populated with thermal and non-thermal radio sources originating during the evolution of short-lived, high-mass stars ($M>8M_{\odot}$), such as H\,{\small{II}}~regions and supernova remnants (SNRs).  The formation of such high-mass stars is controlled by the physical and chemical conditions in the interstellar medium (ISM) that surrounds them, which is affected by these objects themselves, creating a complex feedback loop that is not yet well understood.  Observations of the ISM shed light on its structure and composition and help us understand our Galaxy better by constraining the theoretical models on the evolution of its contents \citep[e.g.,][]{2020A&A...642A.163S,2021A&A...655A.111K,2023A&A...670A..98Y}.  In the past, a plethora of surveys, from sub-millimetre to metre wavelengths, have 
studied objects arising from various stages of stellar evolution, probing angular scales ranging from a few arcseconds to several degrees.  Surveys in the radio regime are particularly helpful as dust is optically thin at low radio frequencies.  The Multi-Array Galactic Plane Imaging Survey \citep[MAGPIS;][]{1994ApJS...91..347B}, the APEX Telescope Large Area Survey of the Galaxy \citep[ATLASGAL;][]{2009A&A...504..415S}, the Coordinated Radio and Infrared Survey for High-Mass Star Formation \citep[CORNISH;][]{2012PASP..124..939H}, the Very Large Array Galactic Plane Survey \citep[VGPS;][]{2006AJ....132.1158S}, the 1--2~GHz HI/OH/Recombination line survey \citep[THOR;][]{2016A&A...595A..32B,2020A&A...634A..83W}, and the 4--8~GHz GLObal view on STAR formation survey \citep[GLOSTAR;][]{2021A&A...651A..85B} are a few such surveys that observed the northern Galactic plane.  

The radio emission at frequencies below 1\,GHz is dominated by nonthermal synchrotron radiation, which is observed not only from SNRs \citep[e.g.,][]{2013tra..book.....W} but also from some star forming regions \citep[e.g.,][]{1999ApJ...513..775W,2019MNRAS.482.4630V}.  The Galactic plane at these frequencies, however, is relatively unexplored at near-arcsecond angular resolution.  The GaLactic and Extragalactic All-sky Murchison Widefield Array survey \citep[GLEAM;][]{2019PASA...36...47H} at 70--230\,MHz covers a large portion of the Galactic plane at $2'$--$4'$ resolution and a sensitivity of about $10-20$~mJy\,beam$^{-1}$.  The 150\,MHz TIFR-GMRT Sky Survey (TGSS) by \citet{2017A&A...598A..78I} covers the Galactic plane at a better resolution of $25''$ and a typical sensitivity of about 5~mJy\,beam$^{-1}$, but the extended emission are poorly recovered due to the imaging choices that were made with regards to their survey goals.  The 843\,MHz Molongolo Galactic plane survey \citep[MGPS;][]{1999ApJS..122..207G} has a sensitivity of about 1--2~mJy\,beam$^{-1}$, a resolution of ${\sim}43''$, and is sensitive to extended structures up to a scale of about $25'$, but it covers only the southern Galactic plane which does not include the first quadrant of the Milky Way.  There are multiple ongoing surveys with the Low Frequency Array below 200~MHz, but they cover only the northern sky (declination $\delta>0\degree$), which does not include the inner Galactic plane \citep[e.g.,][]{2015A&A...582A.123H,2017A&A...598A.104S,2021A&A...648A.104D}.

\begin{figure}
    \centering 
    \includegraphics[clip,trim=4mm 6mm 0 12mm,width=0.47\textwidth]{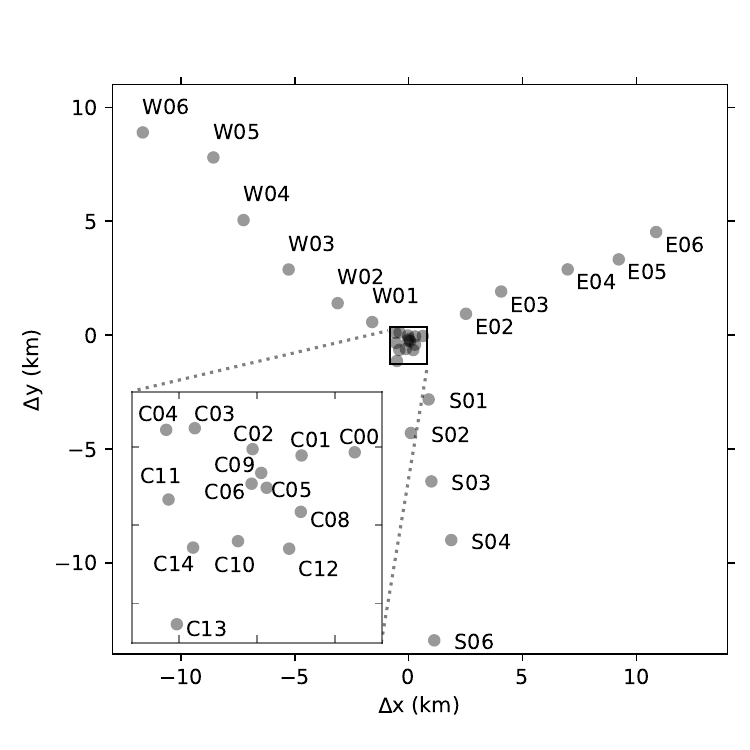}
    \caption{Antenna configuration of the uGMRT array.  The inset shows the central square ($1.6\,\mathrm{km} \times 1.6\,\mathrm{km}$) antennas.  North is upwards and East is rightwards. }
    \label{fig:gmrt}
\end{figure}

The low-frequency regime ($<1$\,GHz) is particularly helpful to study nonthermal emission from SNRs, since they typically have a spectral index of $\alpha{\sim}-0.5$\footnote{$\alpha$ is the spectral index defined using a power-law relation between the flux density, $S_\nu$, and the frequency, $\nu$: ~$S_\nu \propto \nu^\alpha$} and are brighter at lower frequencies.  The number of SNRs identified in the Milky Way so far is lower than 400 \citep{2012AdSpR..49.1313F,2019JApA...40...36G}, which is substantially fewer than the predicted value of ${\sim}1000$--$2000$ \citep{1991ApJ...378...93L,2022ApJ...940...63R}.  Recently, over 150 objects in the first quadrant of the Galactic plane have been identified as SNR candidates \citep{2017A&A...605A..58A,2021A&A...651A..86D}, using the data from the THOR and the GLOSTAR surveys.  Given that most of these candidates are of small angular size \citep{2023A&A...671A.145D} and are expected to have negative spectral indices, typically around $-0.5$,
the next natural step is to search for these objects in the images of high resolution Galactic plane surveys at sub-GHz frequencies.  In addition, due to synchrotron-ageing in old SNRs, the radio spectrum becomes steeper as the higher energy photons are lost much more easily than the lower energy photons.  At frequencies below 100\,MHz, synchrotron self-absorption effects become prominent and make it harder to study nonthermal emission.  Some SNRs are known to have turnover frequencies even close to 1\,GHz or higher \citep{2011A&A...536A..83S,2020MNRAS.496..723K,2023A&A...671A.145D}.  This makes the low-frequency radio window the most suitable to observe and study SNRs.  The THOR and the GLOSTAR surveys have also been used to study the warm-ionized medium \citep{2019ApJ...887L...7S}, cloud formation processes \citep{2020A&A...634A.139W}, stellar feedback \citep{2019A&A...622A..48R}, OH and methanol masers \citep{2019A&A...628A..90B,2021A&A...651A..87O,2022A&A...666A..59N}, and the Galactic center star-formation rate \citep{2021A&A...651A..88N}, to name a few.  However, these centimetre wavelength surveys of the Milky Way lack an equivalent survey below 1\,GHz (at wavelengths longer than 30 cm) that is sensitive to large-scale emission at arcsecond-level resolution covering the first quadrant of the Galactic plane.  

The Giant Metrewave Radio Telescope (GMRT) is an excellent instrument for this purpose.  With its recent upgrade to wide-band receivers and correlator backend \citep[called uGMRT from now on;][]{2017CSci..113..707G}, which can provide an instantaneous bandwidth up to 400~MHz, a sensitive survey is now feasible within a reasonable time-frame.  The expected sizes of the synthesized beams of uGMRT observations conducted in the band-3 (300--500~MHz) and the band-4 (550--750~MHz) are about $5''$ and $8''$, respectively.  Due to its unique hybrid configuration (see Fig.\,\ref{fig:gmrt}), extended structures up to scales of ${\sim}20'$ (at 400\,MHz) can also be recovered.  Inferring the dominant emission mechanism of an object will be possible by comparing the data from the two frequency bands (or even at different frequencies within one band), 
and one will be able to draw broad-band spectral energy distributions for objects with counterparts in the THOR and the GLOSTAR surveys.

To understand the feasibility of such a large-scale survey, we conducted uGMRT observations of the W43/W44 region ($29\degree<l<35\degree$ and $|b|<1\degree$) using the band-3 and band-4 receivers as a part of a `pilot' study.  The low-frequency radio regime (below 1\,GHz) is particularly susceptible to man-made radio frequency interference (RFI) from consumer electronics and artificial satellites.  Moreover, imaging structures larger than a few arcminutes with interferometers, at arcsecond resolution and poor $uv$-coverage (due to snapshot observations), is quite challenging.  This problem gets worse at frequencies below 1\,GHz, where the ionospheric conditions greatly affect the phase stability, due to which achieving a good positional accuracy is difficult.  The chosen region of the Galactic plane for this pilot study covers a region where the Scutum–Centaurus spiral arm meets the bar of the Milky Way.  It is home to several bright extended Galactic sources such as the W44 SNR (G34.7$-$0.4) and the W43 star-forming ``mini-starburst'' region (at $l{\sim}30\degree$, $b{\sim}0\degree$), and contains other compact and ultracompact H\,{\small{II}}~regions, in addition to numerous unresolved extragalactic sources \citep{2020MNRAS.492.2236C}.  This region has not yet been explored with the uGMRT.  All this together makes it an excellent test-bed for our purpose.  The rest of the paper is organized as follows:  In \S\ref{sec:data}, we describe the observations and the data reduction steps.  In \S\ref{sec:results}, we present the continuum and spectral index mosaics, along with the results of the analysis of positions and flux densities.  Finally, the conclusions from this study and its implications for future low frequency Galactic plane surveys are discussed in \S\ref{sec:conclusions}.

\section{Data}
\label{sec:data}

\subsection{Observations}

The data presented in this paper were obtained from observations with the uGMRT that were carried out under the proposal code $36\_061$.  The total time allotted to the project was 26 hours.  We observed in multiple observing sessions in 2019 from May to August (see Table~\ref{tab:sessions}).  We were unable to recover any useful data from the June (band-3) observation, during which we used a 100~MHz bandwidth, due to strong broad-band RFI from artificial satellites near the target sky positions.  This prompted us to re-observe in band-3 with a larger bandwidth of 200\,MHz to account for the sensitivity loss due to channels corrupted by RFI.  These observations were done in August 2019.  

\begin{figure}
    \centering 
    \includegraphics[width=0.49\textwidth]{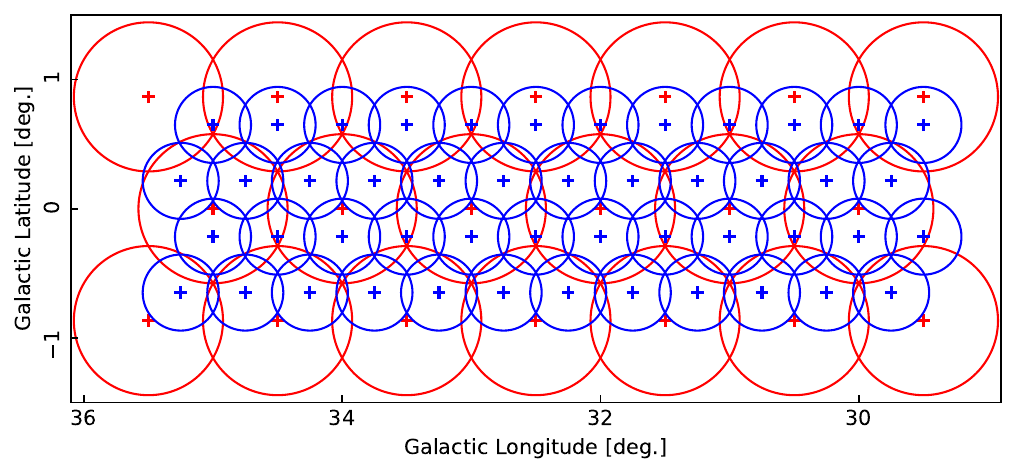}
    \caption{Pointing configuration map of our pilot survey.  Each large red circle marks the field of view of the pointing used for making a mosaic of the band-3 data, while the red cross shows the pointing center.  Similarly, the band-4 fields of view and pointing centers are marked in blue.  }
    \label{fig:pointings}
\end{figure}

The observed region covers the Galactic longitude range $l=29\degree-35\degree$, from the Galactic latitude $b=-1\degree$ to $b=+1\degree$.  We followed a hexagonal mapping pattern to observe the target region in several individual pointings.  The pointing centers are shown in Fig.~\ref{fig:pointings}, along with the fields of view used for making a mosaic of the images.  The fields of view that were chosen for each pointing in band-3 and band-4 are about $35'$ and $18'$, respectively, which correspond to primary beam attenuation factors of 0.36 and 0.64, respectively.  Each target field was observed for ${\sim}$14 minutes in total, done in two separate scans of seven minutes each that were spaced across the observation time in order to achieve a slightly better $uv$-coverage.  To establish the flux density scale, the standard primary calibrator 3C\,286 was observed at least once during an observation, and the gain calibrator J1851+005 was observed about once every hour.  The best continuum point source sensitivity achievable is expected to be about 150\,$\micro$Jy\,beam$^{-1}$ and 60\,$\micro$Jy\,beam$^{-1}$ in band-3 and band-4, respectively, with our integration time and usable bandwidth, based on the GMRT exposure time calculator\footnote{\url{http://www.ncra.tifr.res.in:8081/~secr-ops/etc/rms/rms.html}}.  The expected resolutions of the data are about $8''$ and $5''$ in band-3 and band-4, respectively.  All uGMRT data are recorded in spectral line mode, typically in 8192 channels, which helps in isolating the RFI to a fraction of the total number of channels across the observed bandwidth.  The steps to reduce the continuum data and produce mosaics are explained below.

\begin{figure}
    \centering 
    \includegraphics[width=0.49\textwidth]{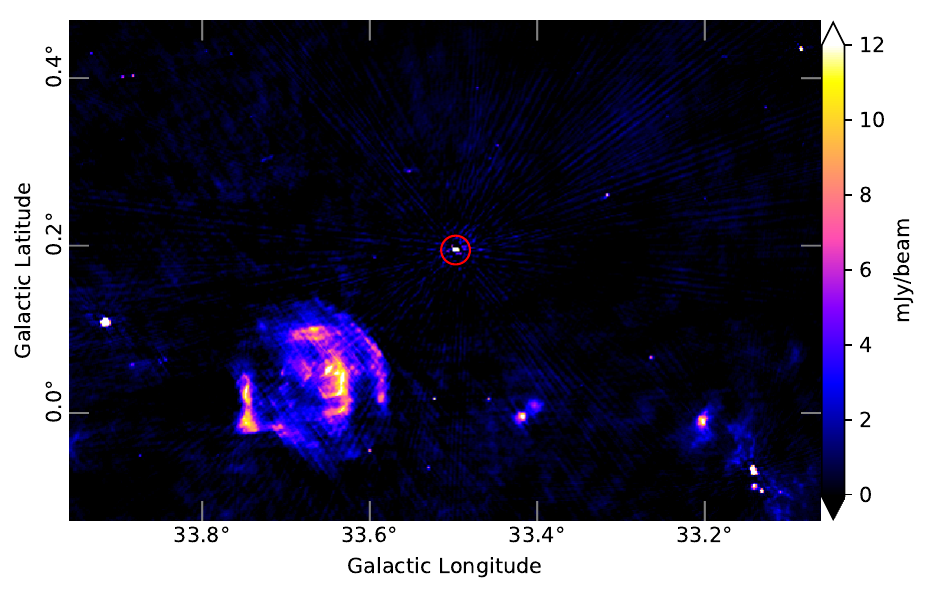}
    \caption{650 MHz image of the gain calibrator field (J1851+005) of our observations.  Circled in red at the center is the phase calibrator (J1851$+$005; Galactic coordinates $l=33.5\degree$, $b=0.2\degree$).  The large and extended structure to its southeast is the supernova remnant (SNR) G33.6$+$0.1.  The beam size is about 10 arcsecond.  The intensity scale is saturated to clearly bring out the weak extended structures. }
    \label{fig:phasefield}
\end{figure}

\begin{table}
\caption{MeGaPluG pilot survey observations}
\label{tab:sessions}
\centering    
\begin{tabular}{c c c c}
\hline\hline
Date & band & Frequency range & Usable data \\
     &      & (MHz) &  (\%) \\
\hline
2019-05-15 & 4 & 550--750 & ${\sim}$75 \\
2019-05-26 & 4 & 550--750 & ${\sim}$66 \\
2019-05-27 & 4 & 550--750 & ${\sim}$57 \\
2019-06-12 & 3 & 350--450 & ${\sim}$5 \\
2019-08-22 & 3 & 300--500 & ${\sim}$53 \\
\hline
\end{tabular}
\end{table}

\subsection{Data reduction}
\label{subsec:datareduction}

\begin{figure*}
    \centering 
    \includegraphics[width=0.8\textwidth]{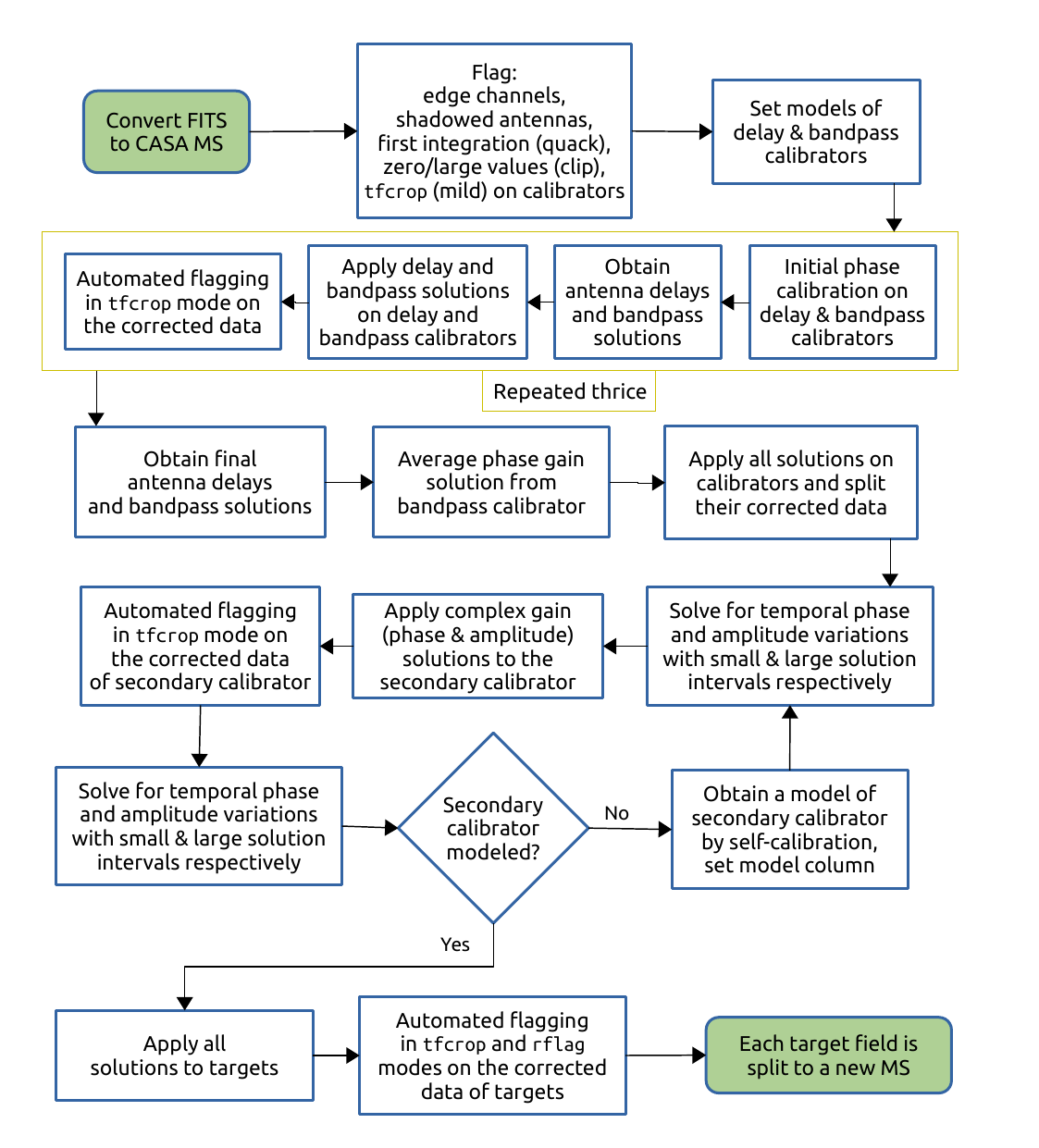}
    \caption{Calibration scheme that we followed to reduce the data.  }
    \label{fig:calflowchart}
\end{figure*}

The raw visibility data measured by an interferometer are corrupted due to variable response from the receiving instruments and the atmosphere (so-called ``gains'').  The process of calibration estimates the actual visibilities from the observed visibilities, which are related to each other by a measurement equation as described by \citet{1996A&AS..117..137H}.  Imaging the calibrated data constitutes the gridding of the visibilities, performing the inverse Fourier transform of the gridded visibilities, and removing the imprint of the point spread function (PSF, the ``dirty beam'') of the interferometer.  These two steps (calibration and imaging) are detailed in the following sections.

\subsubsection{Calibration}
\label{subsubsec:datacalib}

The data were calibrated in the CASA software suite \citep{2007ASPC..376..127M}\footnote{\url{https://casa.nrao.edu/}}.  The compact quasar 3C\,286 served as the primary flux density and bandpass calibrator with its flux density scale given by \citet{2017ApJS..230....7P}, and J1851+005 was used as the complex gain calibrator.  A routine direction-independent calibration strategy was followed, except for the gain calibrator field.  As J1851+005 lies in the Galactic plane with bright and extended sources in its vicinity (see Fig.\,\ref{fig:phasefield}), we first generated a model of this region by self-calibrating this field and then obtained new gain calibration solutions from the model, which were later applied to the target fields. 
Data corrupted due to RFI were ``flagged'' (i.e., removed from further use) using automated techniques such as the \texttt{tfcrop} and the \texttt{rflag} modes of the CASA task \texttt{flagdata}.  We used some strategies and modules of the VLA scripted pipeline\footnote{\url{https://science.nrao.edu/facilities/vla/data-processing/pipeline/scripted-pipeline}} to make our calibration scripts fast and reliable.  Due to the bright extended structures in the Galactic plane that we intend to detect and study, direction-dependent calibration is complicated in this case \citep[e.g.,][]{2020A&A...635A.147A}.  Even with full synthesis observations, imaging extended structures with the GMRT using direction-dependent gains is challenging \citep[see][for example]{2014MNRAS.442.2867W}, and we do not find any previous studies that have successfully imaged large and bright extended sources with direction-dependent calibration.  Hence we decided to proceed with only the standard direction-independent calibration.  A flowchart of the calibration process is shown in Fig.\,\ref{fig:calflowchart}, and a full description is given in the Appendix \ref{apdx:calscheme}.

Preliminary images of the calibrators were produced with the task {\tt tclean} in order to confirm that the calibration procedure is successful.  The task {\tt statwt} was applied to the target data to weigh down outliers based on local variances. This was helpful to reduce the imaging artefacts caused by residual RFI which may not have been flagged during the previous steps.  The corrected data of the target area are now considered ready for further processing.

\begin{figure*}[ht]
    \centering
    \includegraphics[width=\textwidth]{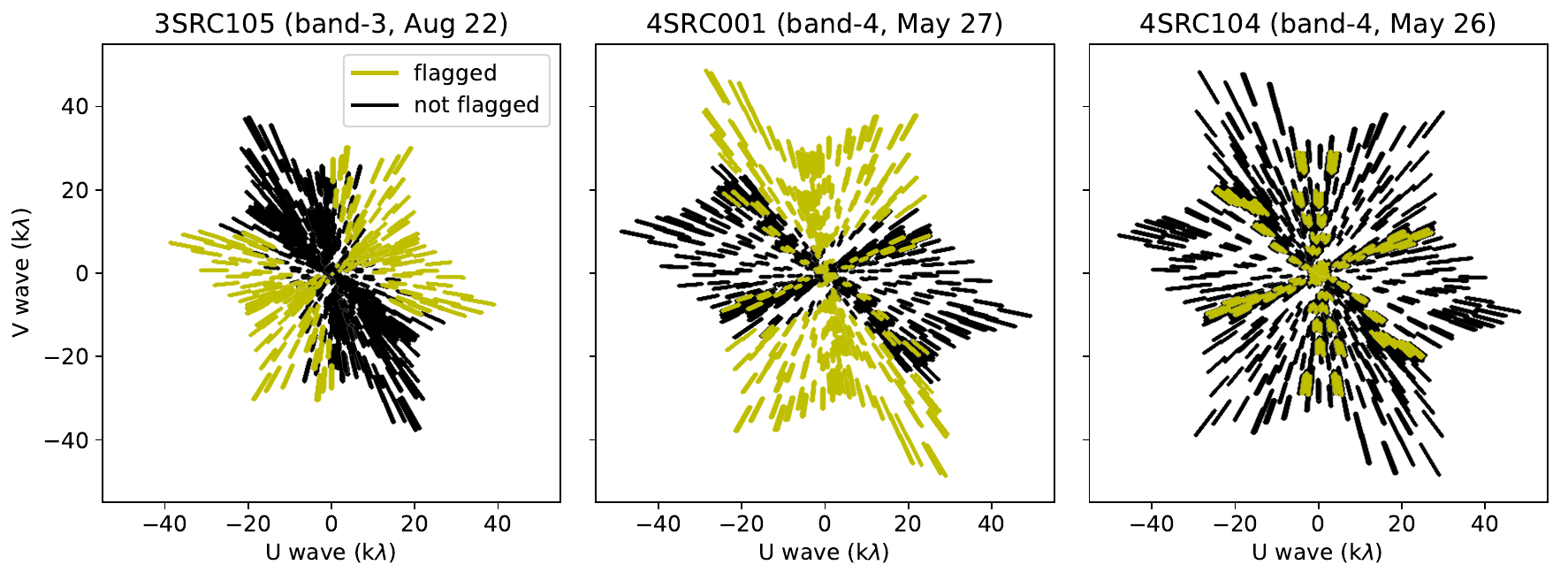}
    \caption{$uv$-coverage plots of typical fields observed on 2019 August\,22 (\textit{left}, 3SRC105), May\,27 (\textit{middle}, 4SRC001), and May\,26 (\textit{right}, 4SRC104).  The asymmetry caused by non-functional antennas (marked in yellow) is clearly visible in the fields observed on August\,22 and May\,27, which leads to an elongation of the synthesized beam.}
    \label{fig:uvcov}
\end{figure*}

At the end of this calibration procedure, about 50\% of the data in band-3 were flagged, which is typical of uGMRT observations taken at  these frequencies due to strong RFI.  In band-4, ${\sim}$15$-$45\% of the data were flagged, depending on the pointing and the date of the observation (see Table~\ref{tab:sessions}).  RFI at these frequencies is relatively less of an issue, and a large fraction of the flagging was caused by non-working antennas.  The band-4 data from May\,27 were the most affected; five antennas were not in working condition during this observation in addition to the intermittent RFI typical at low frequencies, leading to a total of ${\sim}$43\% data loss (see Fig.~\ref{fig:uvcov}).  We also find that several fields had a number of central square antennas flagged for a portion of the scans during the calibration due to RFI, especially in band-4.  This meant that the large scale structure is not well recovered for these fields, and the largest angular scales to which the data are sensitive are typically about ${\sim}6'$ and ${\sim}4'$ in band-3 and band-4, respectively, which are significantly smaller than the expected values of ${\sim}25'$ and ${\sim}15'$.

\subsubsection{Polarization}
\label{subsubsec:polarization}

The band-4 observations (550$-$750 MHz) were done in full-polarization mode, which allows one to make images of the Stokes $Q$, $U$, and $V$ data in addition to Stokes $I$.  Linearly polarized emission is expected from synchrotron emitters such as SNRs whereas emission from H\,{\small{II}}~regions due to thermal Bremmstrahlung is unpolarized. Hence polarization information plays a key role in separating thermal and nonthermal emission, and thus in identifying new SNRs \citep[e.g.,][]{2021A&A...651A..86D}.  The polarization calibration involves obtaining delays of cross-hand polarizations (RL, LR), characterizing instrumental polarization by determining the frequency-dependent leakage terms (colloquially known as the `D-terms'), and calibrating the absolute polarization position angle using a well-known polarized calibrator.  The D-terms can be estimated either by observing an unpolarized source for a short interval, or by tracking a polarized calibrator with a good coverage of parallactic angle.  In order to minimize the observation time, we chose to observe 3C\,48, which is unpolarized at these frequencies \citep{2013ApJS..206...16P,2014MNRAS.437.3236F}, using which we were able to determine the leakage terms.  For calibrating the polarization position angle, we used 3C\,286 following \citet{2019MNRAS.487.4819M}.  However, we obtained a polarization fraction of ${\sim}3.5$\% for 3C\,286 at 650\,MHz, which agrees with the value of $2.7$\% at 610\,MHz by \citet{2012PhDT.........4F} but is in tension with the value of $7.6$\% at 607\,MHz given by \citet{2019MNRAS.487.4819M}.  We note that the polarization properties of 3C\,286 are not well known below 1\,GHz, and it is observed to strongly depolarize (to lower than $1$\%) at frequencies below 500~MHz \citep{EVLAmemo207}.  We imaged the Stokes $Q$ and $U$ data for a field with the W44 supernova remnant, where we expected a polarization signal of at least a few percent, but we were unable to recover any.

Sources with an absolute value of the rotation measure (RM)\footnote{RM is defined using $\Delta \chi = \mathrm{RM} \cdot \lambda^2$, where $\Delta \chi$ is the change in the electric vector position angle and $\lambda$ is the wavelength.} higher than ${\sim}7~\mathrm{rad\,m}^{-2}$ would appear depolarized in our images, since their polarization angle rotates multiple times over the 200~MHz bandwidth.  Galactic sources can have RMs much larger than ${\sim}7~\mathrm{rad\,m}^{-2}$, and RMs may even be in the range of hundreds of rad\,m$^{-2}$ \citep[e.g.,][]{2000ApJ...542..380G}.  To be able to detect the linear polarization from sources with RMs as high as ${\sim}100~\mathrm{rad\,m}^{-2}$, we split the data into 20 frequency bins, imaged each bin and made the linearly polarized intensity image of each bin separately, and finally combined them by averaging to form a single output image.  However, we were unable to obtain any useful Stokes $Q$ and $U$ images, which is due to the poor signal-to-noise ratio in each bin.  This is not unexpected, since the total time spent on each pointing is less than 15 minutes and the linearly polarized intensity is expected to be only a fraction of the total intensity, meaning the image reconstruction in each frequency bin will be quite poor.  Hence, we imaged only the Stokes $I$ data.

\subsubsection{Imaging}
\label{subsubsec:imaging}

Generally, imaging is the most intensive part of the reduction of radio interferometric data, both in terms of CPU time and manual effort.  The task {\tt tclean} of CASA was employed for imaging.  Several deconvolution algorithms based on the Cotton-Schwab CLEAN are available in this task.  It was used in parallel mode on four cores in order to speed up the imaging process.  The complex structures in the target region, the wide bandwidth of the receivers, and the wide field of view at low frequencies pose significant challenges in imaging.  We briefly explain our imaging strategies below:

\begin{itemize}

\item{\it w-projection:}
To account for the non-coplanar baseline effect, we used w-projection \citep{2008ISTSP...2..647C}.  The number of planes was set to be automatically calculated using the parameter \texttt{wprojplanes=-1}.

\item{\it Outliers:}
We imaged a region that extends at least out to the first null of the primary beam (${\sim}2.3\degree$ and ${\sim}1.4\degree$ at 400~MHz and 650~MHz, respectively), and in some cases we used outlier fields as well.  This was done in order to image very bright sources that are located well outside the primary beam.  Such sources, if not CLEANed, can produce noticeable features and significantly increase the root mean square (rms) background noise.

\item{\it Wide bandwidth:}
In order to account for the 
frequency-dependent variations across the 200 MHz bandwidth, we used the Multi-Term Multi-Frequency synthesis (MT-MFS) algorithm \citep[\texttt{specmode='mfs'} and \texttt{deconvolver='mtmfs'};][]{2009IEEEP..97.1472R,2011A&A...532A..71R} with two Taylor coefficients in the spectral model (\texttt{nterms=2}).

\item{\it Multiple scales:}
To model emission structures of various sizes, we used the multi-scale CLEAN \citep{2008ISTSP...2..793C} along with the aforementioned MT-MFS algorithm.  The scales were chosen such that they roughly correspond to the sizes of structures expected to be detected in our observations of a given region.

\item{\it Thresholds:}
The MT-MFS algorithm in the task \texttt{tclean} is prone to divergence if the stopping criteria and the scales are not carefully set, especially if the noise threshold to be reached is set to a value that is lower than the achievable noise levels.  Hence, we used a dynamic threshold based on the local image statistics, fixed the maximum number of minor cycle iterations to 2000 (with the parameter \texttt{cycleniter}), and lowered the loop gain from the default of 0.1 to 0.06.  This came at the cost of a larger computational footprint and more processing time, due to major cycles being visited more often.  However, we found that the algorithm does not diverge with these parameters, and the images were also more accurate due to better residual image construction.

\item{\it Weighting:}
The Briggs weighting scheme \citep{1995AAS...18711202B} with a robust parameter of 0.5 was used, which is a compromise between the best achievable resolution and the best sensitivity to large-scale structures.

\item{\it Pixel size:}
The pixel size was chosen such that there are at least three pixels across the minor axis of the expected synthesized beam.  This 
allows a good Gaussian fit for the central lobe of the point spread function.

\item{\it Masking:}
It is impractical to manually identify real emission for CLEAN masks since we covered a large region.  Therefore, we made use of the auto-masking feature of \texttt{tclean} \citep[\texttt{usemask='auto-multithresh'};][]{2020PASP..132b4505K}.  This also helps to keep the imaging process reproducible and minimize human errors and bias. 

\item{\it Self-calibration:}
By modelling the phases from an initial shallow CLEAN, we self-calibrated the target fields, and repeated this process twice to generate a better model and a better final image.  Although the improvement in dynamic range was quite low (${\sim}15\%$) -- which was expected due to the poor $uv$-coverage -- we found that it improved the image fidelity by reducing the sidelobe artefacts caused by imperfect deconvolution.

\item{\it Wide-band primary beam correction:}
The attenuation caused due to the primary beam response was corrected for using the contributed CASA task \texttt{wbpbgmrt}\footnote{\url{https://github.com/ruta-k/uGMRTprimarybeam}}.

\item{\it Uniform resolution:}
Due to the time-variable RFI and the non-working antennas, the resolution achieved greatly depended on the field and the time of
an observation.  For the sake of uniformity and to allow comparison of  the images of both bands, we restricted the $uv$-range to $10k\lambda$ to make images with a common circular beam.  The `full-resolution' images are made separately with no restriction on the $uv$-range.

\item{\it Mosaics:}
We imaged each field individually first and combined them later to form a mosaic using the tool \texttt{linearmosaic} of CASA, where the pixels in the overlapping regions are weighted by using the primary beam response as the sensitivity.  We note that while wide-band mosaic imaging is available in CASA\footnote{\url{https://casadocs.readthedocs.io/en/stable/notebooks/synthesis_imaging.html\#Wide-Band-Imaging}}, the vastly varying beam sizes across the target fields observed during different nights (see \S\ref{sec:results}) make it unsuitable for imaging our data.  

\item{\it Coordinate transformation:}
Finally, the mosaic was re-gridded from equatorial to Galactic coordinates using the software Montage \citep{2003ASPC..295..343B}. 

\end{itemize}

\begin{figure}
    \centering 
    \includegraphics[width=0.49\textwidth]{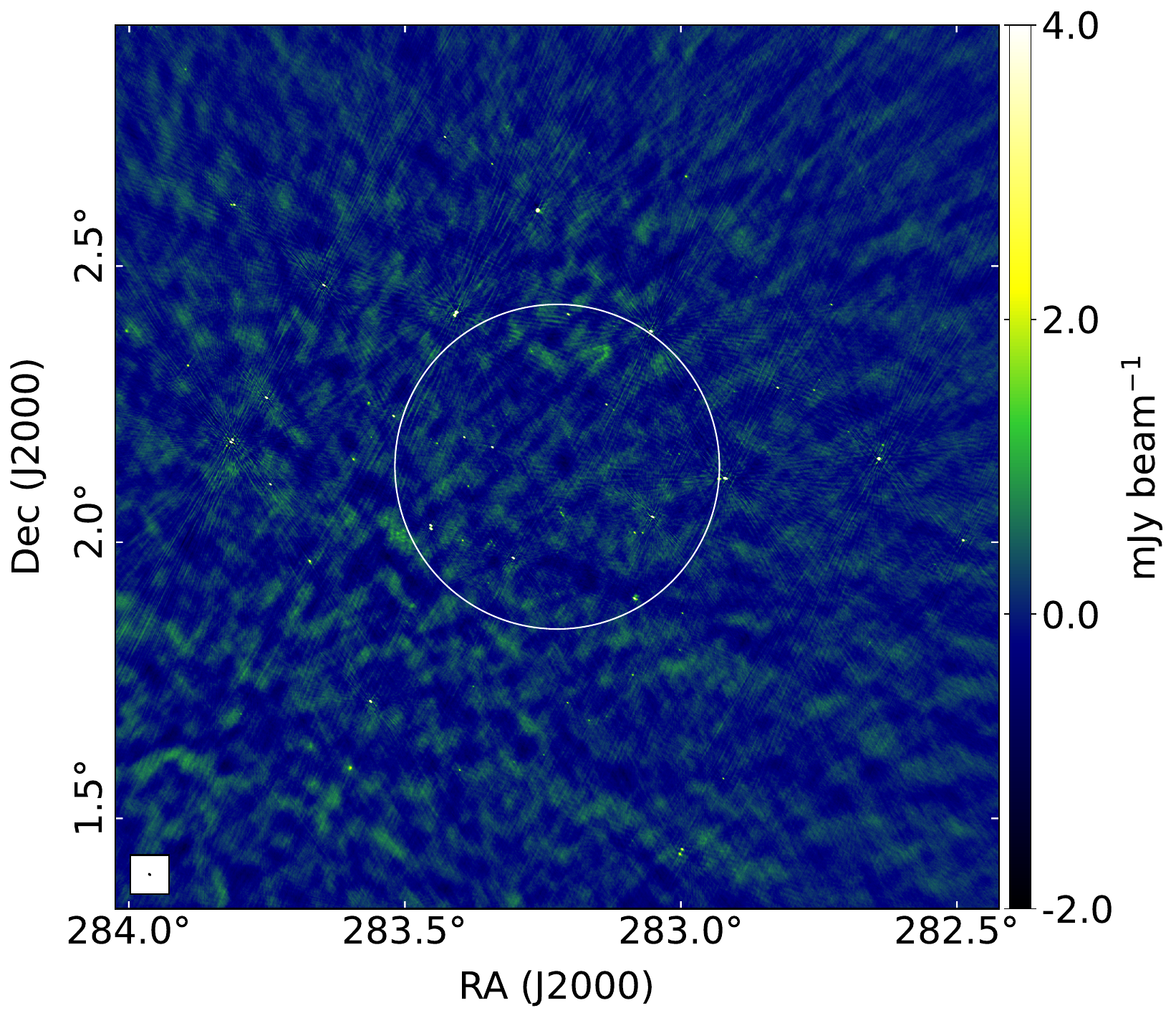}
    \caption{Example field observed in band-4 during a night with all antennas functioning, uncorrected for primary beam attenuation, in the native equatorial coordinates.  The white circle marks the region used for mosaicking.  The beam size is $12'' \times 7''$, and the rms noise is about 0.3 mJy\,beam$^{-1}$.  The colormap and the limits are chosen such that the low-level noise features and the negative sidelobes are seen clearly.  }
    \label{fig:examplefullres}
\end{figure}

We found that the synthesized beams obtained after imaging depended strongly on the day of the observation (due to the varying number of working antennas and the time-variable RFI) and also on the software being used.  AIPS, and the versions of CASA prior to v5.8, have a PSF-fitting algorithm that works reliably only if the range of over-sampling factor (number of pixels across an axis of the expected synthesized beam) is from three to five.  Because of the non-functional outer antennas during the  May 27 and August 22 observations (see Fig.\,\ref{fig:uvcov}), employing 
an over-sampling factor of $3-5$ was not simultaneously possible on both the major and minor axes of the synthesized beam when we produced images  using visibilities from all functional baselines.  This led to an incorrect beam size estimation if AIPS or an older version of CASA was used for imaging (see Appendix \ref{apdx:beamsize} for more details).  Hence, we used a version of CASA, v5.8, in which the PSF fitting algorithm was updated\footnote{\url{https://casadocs.readthedocs.io/en/v6.2.0/notebooks/introduction.html}} that is able to deliver correct results even in extreme cases such as ours.  Unfortunately, the non-working antennas meant that the final common beam size for both bands we were able to obtain is $25''$, which is about three times larger than the best possible resolution of about $8''$.  In Fig.\,\ref{fig:examplefullres}, we show an image resulting from a  pointing observed in band-4 on a night with all working antennas, with no primary beam correction and at its native resolution.  An rms noise of about 0.3 mJy\,beam$^{-1}$ and a beam-size of $12'' \times 7''$ were achieved, which are close to the theoretical values for this observation.

Background noise maps corresponding to each band were created using the software {\tt SExtractor} \citep{1996A&AS..117..393B}.  A spectral index map was created using the images from the two bands, after reprojecting them onto the same grid.  This map was obtained on a pixel-by-pixel basis, and the calculation was done only if the signal-to-noise ratio of a pixel in both images was greater than three.  We note that although the multi-frequency synthesis (MFS) algorithm in CASA \citep{2009IEEEP..97.1472R} is capable of producing spectral index maps directly during imaging, we find it to deliver un-physical spectral indices for our observations.  We routinely obtained spectral index values of the order of tens even after accounting for the primary beam attenuation.  This was true not just for faint or extended emission, but also for point-like sources with signal-to-noise ratios about 50.  A recent study by Rashid et al. (submitted) analyzed this issue using CASA simulations and also came to the same conclusion that the spectral index maps produced by the MFS algorithm are not reliable for uGMRT data with a signal-to-noise ratio below 100.  Hence, for this work, we use only the spectral index map derived using the band-3 and the band-4 mosaics together.

\subsection{Missing flux density}
\label{subsec:missFD}

All interferometers spatially filter out a fraction of the flux density of extended sources.  The angular scale at which such filtering starts to affect the flux densities depends on the $uv$-coverage and the structure of the extended source, but an upper limit of the largest angular scale recovered in a full-synthesis observation ($\theta_\mathrm{LAS}$) can be estimated using
\begin{equation*}
    \theta_\mathrm{LAS} \lesssim \frac{\lambda}{b_\mathrm{min}},
\end{equation*}
where $\lambda$ is the wavelength and $b_\mathrm{min}$ is the length of the shortest available baseline (${\sim}$100\,m for the uGMRT).  We found that the values of $\theta_\mathrm{LAS}$ in our observations are about $6'$ and $4'$ in band-3 and band-4, respectively, and in some regions it is even smaller.  These are significantly lower than the estimates for two reasons.  One, we observed in the snapshot-mode, implying that the $uv$-coverage is quite poor compared to a full-synthesis observation.  Two, the RFI situation at the observatory is such that it affects the central square antennas much more than the outer antennas.  This further degrades the quality of image reconstruction of extended structures.  In order to mitigate this issue of missing flux density, zero-spacing information can be added from a dataset at the same frequency that recovers the extended emission, such as the data from a single-dish telescope.  This can be done using a method called `feathering', or by joint deconvolution \citep[e.g.,][]{1984ApJ...283..655V,2011ApJS..193...19K,2019AJ....158....3R,2023A&A...671A.145D}.  In any case, the diameter of the single-dish telescope must be greater than the shortest baseline length for the combination to work well, since, only then an overlap in the $uv$-coverage is guaranteed.  Only two steerable single-dish telescopes exist that are suitable for such a zero-spacing correction of the GMRT data: the 100\,m Effelsberg telescope and the 100\,m Robert C.~Byrd Green Bank Telescope.  While there are receivers on both the telescopes that can complement the frequency coverage in our study, the RFI from man-made electronics at both the observatories renders a large fraction of the bandwidth useless.  It currently remains unfeasible to perform a large-scale survey using the available bandwidth on these receivers to reach the sensitivity required to be added to our uGMRT data.

\begin{figure*}
  \centering
  \includegraphics[width=0.89\textwidth]{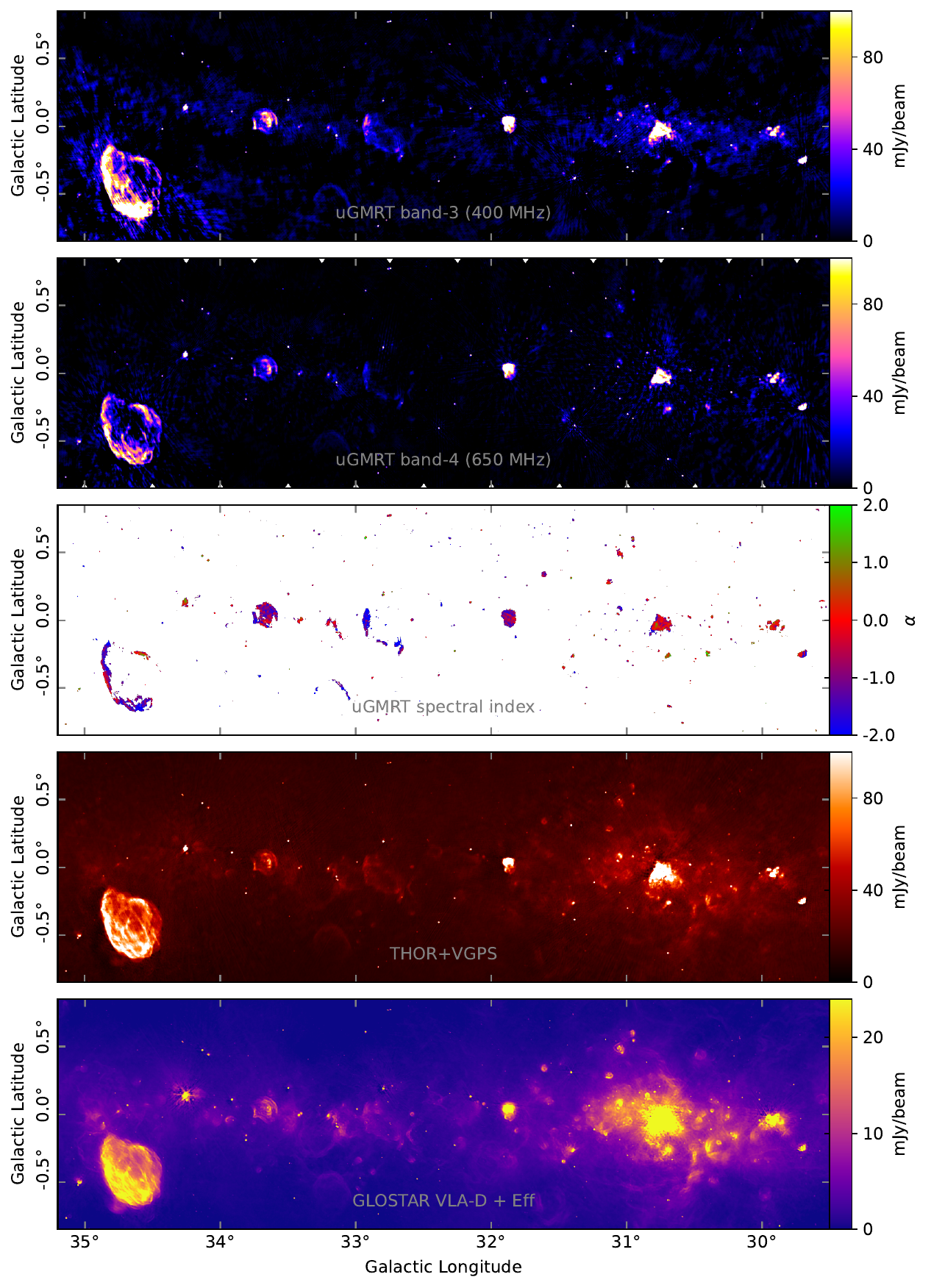}
  \caption{From top to bottom: band-3 and band-4 mosaics from this project convolved to the common beam size of $25''$, their resulting spectral index map, the 1.4 GHz THOR image of the same area combined with the VGPS data \citep{2016A&A...595A..32B,2006AJ....132.1158S}, which is also at $25''$ resolution, and the 5.8~GHz GLOSTAR data (VLA D-configuration image combined with the Effelsberg image), which is at a resolution of $18''$.}
  \label{fig:b3b4mos}
\end{figure*}

\begin{figure*}
    \centering 
    \includegraphics[width=\textwidth]{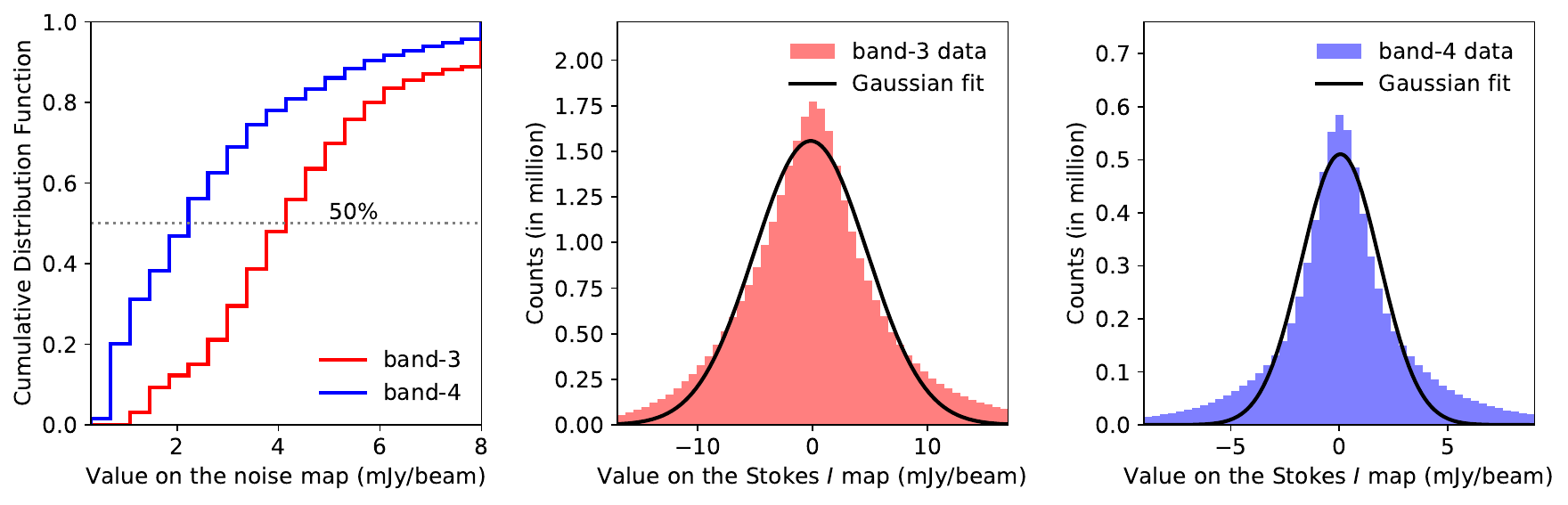}
    \caption{Cumulative distribution functions of the noise maps (\textit{left}), and the histograms of values of the pixels in the continuum mosaics of band-3 (\textit{middle}) and band-4 (\textit{right}).  Gaussian least-squares fits are performed on the histograms, which gave a standard deviations as ${\sim}4.9$ mJy beam$^{-1}$ for band-3 and ${\sim}1.8$ mJy beam$^{-1}$ for band-4. }
    \label{fig:Ihist}
\end{figure*}

\begin{figure*}
    \centering 
    \includegraphics[width=\textwidth]{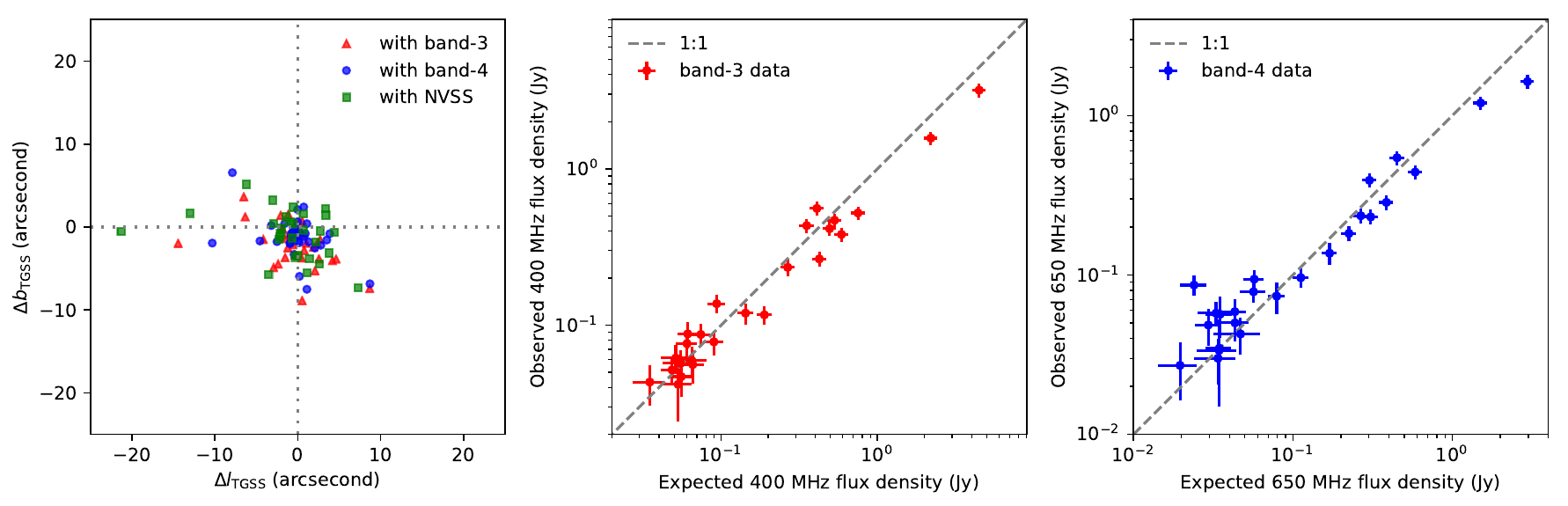}
    \caption{\textit{Left}: Position offsets of the TGSS sources seen in our survey and the NVSS, with obvious extended sources excluded.  Observed and expected flux density comparison for sources in band-3 (\textit{middle}) and band-4 (\textit{right}), where the expected flux density is obtained using spectral indices from \citet{2018MNRAS.474.5008D}.  }
    \label{fig:locFDcomp}
\end{figure*}

\section{Results}
\label{sec:results}

The mosaics of band-3 and band-4 at a resolution of $25''$, along with the spectral index map, are shown in Fig.\,\ref{fig:b3b4mos}.  All the uGMRT images in the figure are dynamic range limited, with rms noise levels of about ${\sim}5\mathrm{~mJy~beam}^{-1}$ in band-3 and ${\sim}2\mathrm{~mJy~beam}^{-1}$ in band-4.  The distributions of the brightness values of the pixels in the 
mosaics and the cumulative distributions of the values of pixels in their respective background noise maps are shown in Fig.\,\ref{fig:Ihist}.  We notice sidelobe artefacts near a few bright point sources in both bands, which are likely due to the lack of direction-dependent calibration and imperfect deconvolution in conjunction with the sparse $uv$-coverage.  Bright sources that are not close to the center of the pointing corrupt the image, since the gains are truly applicable only at the phase center of the field.  In addition, the regions close to the SNR W44 (G34.7$-$0.4) have a larger rms noise because the SNR is quite bright and extended with a complicated structure.  In the band-3 mosaic, in the region enclosed between the Galactic coordinates $l=34\degree$--$35\degree$ and $b=0\degree$--$0.5\degree$, there appears to be a very faint, large-scale structure bending away from the Galactic mid-plane.  This feature, however, does not have a counterpart in the 70--230\,MHz GLEAM images \citep{2019PASA...36...47H} or the THOR+VGPS data \citep{2016A&A...595A..32B}, and its surface brightness is close to the local noise level as well.  So, this feature is likely an artefact of the imaging process.

\subsection{Comparison with other surveys}

To date, there is no survey at high angular resolution that covers the region surveyed in our pilot study at frequencies of 300--750~MHz.  Hence we use data from surveys conducted at other frequencies to compare the positions and flux densities of the sources detected in our survey.  

The 150 MHz TIFR-GMRT sky survey \citep[TGSS;][]{2017A&A...598A..78I} covers the region we observed at the same resolution ($25''$) and a comparable sensitivity (${\sim}10\mathrm{~mJy~beam}^{-1}$ in the Galactic plane).  Since the inner baselines were severely down-weighted during the data reduction, only the brightest edges of extended sources were detected in their images.  The catalog of this survey lists 50 sources in the region we observed.  We detect all the sources except one, named J185141.6+003739 in the TGSS catalog.  Its signal-to-noise ratio is about nine, but this source is not detected in any radio survey other than the TGSS, according to the SIMBAD database.  There are two bright sources very close to this object: J185146.7+003531, which is in fact our gain calibrator, and the SNR G33.6+0.1, which is an extended source (Fig.\,\ref{fig:phasefield}) and thus not well sampled in the TGSS images.  Hence, the sole TGSS source that we do not detect in our data is likely to be a `ghost' \citep{2014MNRAS.439.4030G}\footnote{Ghost sources are calibration artefacts but show a PSF similar to the instrumental PSF.  They arise due to imperfect sky models and direction-dependent effects.}.  Of the remaining 49 TGSS sources, 17 are clearly either a part of a supernova remnant or are double-lobed sources that are almost certainly distant radio galaxies, and the other 32 are compact sources.

The NRAO-VLA Sky Survey \citep[NVSS;][]{1998AJ....115.1693C} is a 1.4 GHz survey of the whole sky north of $\delta=-40\degree$, at a resolution of $45''$.  While the NVSS catalog contains over 1300 sources in the area we covered, a visual inspection reveals that most of these sources are obviously false.   This was noted by earlier studies as well \citep{2016A&A...588A..97B}, and this is not unexpected.  The techniques used for data reduction and cataloging of the NVSS data are best suited for imaging and detecting compact extragalactic objects, and do not perform well in very crowded and complex fields such as ours.  However, of the 32 compact sources detected in TGSS, 28 were detected in the NVSS as well, using which we made position and flux density comparisons to validate our mosaics.  For these sources, we evaluated their peak intensities and integrated flux densities, along with their positions, by performing two-dimensional Gaussian fits on our images.  We compared these values to the values from the TGSS and the NVSS source catalogs, and the results are shown in Fig.~\ref{fig:locFDcomp}.  

While the position accuracies of both TGSS and NVSS are reported to be better than $2''$ \citep{1998AJ....115.1693C,2017A&A...598A..78I}, we find that there are significant offsets between the peak positions of a few sources in this field (Fig.\,\ref{fig:locFDcomp}; left panel).  The complex background in the region in question moves the peak of a source along the local noise gradient, which makes many positions listed in these source catalogs less reliable compared to the regions outside the Galactic plane.  In addition to background contamination, the emission at different frequencies may be dominated by different mechanisms, which may result in shifting peak positions, or they may be entirely different sources altogether.  While the ionosphere is expected to contribute to propagation delay at these frequencies, each individual antenna of the GMRT sees a roughly constant ionosphere above ${\sim}400$~MHz.  Hence, antenna-based phase-only self-calibration is usually sufficient \citep{2005ASPC..345..399L,2009A&A...501.1185I} to deal with the resulting phase errors.  Since we have not performed direction-dependent calibration, calibration issues may also play a role in shifting the peak positions.  However, we note that most of the offsets are within 1-2 pixels, and all are smaller than the beam size.

Assuming that a single power-law index holds between 150 MHz and 1.4 GHz and that sources are not variable, a spectral index catalog was prepared by \citet{2018MNRAS.474.5008D} using the data from the TGSS and the NVSS surveys.  It has already been shown to be useful to discriminate between thermal and nonthermal emission from and H\,{\small{II}}~regions and SNRs  \citep{2018ApJ...866...61D}, in addition to pinpointing extragalactic sources.  Here we use this spectral index catalog to derive an expected flux density of the sources at 400 MHz and 650 MHz.  We compare these values with the integrated flux densities from the Gaussian fits to band-3 and band-4 sources, and the results are shown in the middle and right panels of Fig.\,\ref{fig:locFDcomp}.  

Although there appears to be a general agreement between the values observed by us and the derived estimations, the number of bright sources that fall below the 1:1 line is larger than the number above the 1:1 line, i.e., the number of bright sources with observed flux density lower than the expected flux density is visibly high.  While this may be caused by physical reasons such as the assumptions of a single power-law spectral index being invalid and/or variability, it may also point to a systematic error within one of the TGSS, NVSS, and our band-3/4 images.  The NVSS flux densities are based on the scale of \citet{1977A&A....61...99B}, while we used the recent \citet{2017ApJS..230....7P} scale.  On the other hand, the TGSS is scaled using the low frequency scale of \citet{2012MNRAS.423L..30S}, which is not based on an absolute and independent standard, but it is still consistent to within 2\% of the other two scales.  \citet{2017arXiv170306635H} finds systematic position-based amplitude offsets on the angular scale of degrees, which can reach  as high as  40\% in some regions (see their Figure\,3).  It had been noted that the higher system temperatures ($T_{\mathrm{sys}}$) in Galactic plane observations may lead to the receivers being pushed into a non-linear regime\footnote{\url{http://www.ncra.tifr.res.in/ncra/gmrt/gmrt-users/galactic-plane}}.  Since our observations were done with the upgraded receivers, this is unlikely to affect the results, and we find that the flux densities of the primary and secondary calibrators match well with the literature values.

\subsection{Nonthermal emission}

\begin{figure}
    \centering 
    \includegraphics[width=0.49\textwidth]{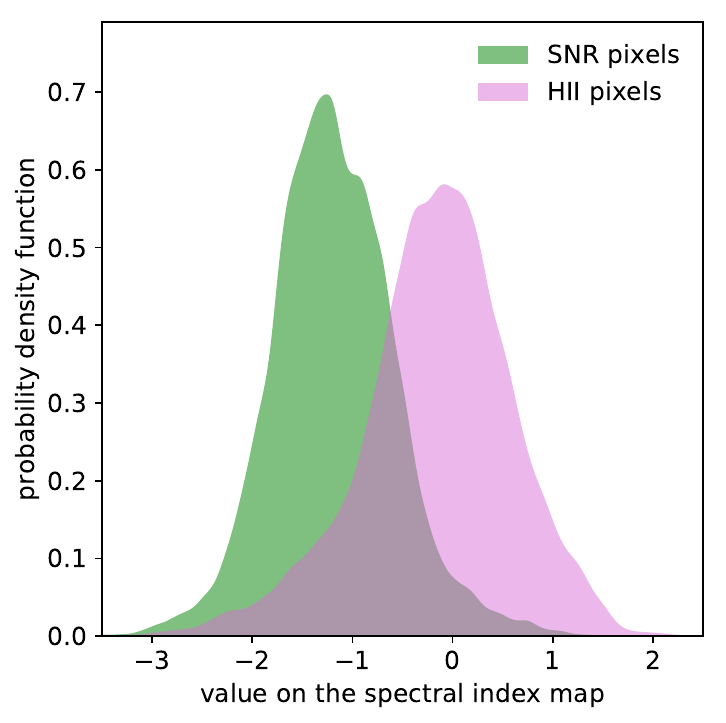}
    \caption{Probability density functions, obtained using kernel density estimations, of the values of spectral index pixels belonging to SNRs and H\,{\small{II}}~regions.  }
    \label{fig:spidx}
\end{figure}

H\,{\small{II}}~regions are routinely misidentified as SNRs in the Galactic plane due to their similar radio morphology \citep[see][for instance]{2021A&A...651A..86D}; however, the spectral index is an excellent discriminant between the two.  H\,{\small{II}}~regions typically have a `flat' or `rising' spectrum at radio wavelengths ($-0.1<\alpha<2$) depending on their optical depth \citep{2013tra..book.....W}. In contrast, shell-type SNRs have $\alpha\lesssim -0.5$ and Crab Nebula-like SNRs (pulsar wind nebulae, PWNe) or composite SNRs have $\alpha\sim -0.2$ \citep{1997ApJ...490..291B,2015A&ARv..23....3D}.  In Fig.\,\ref{fig:spidx}, we show the probability distribution functions of the pixels in the spectral index map within the angular extents of known SNRs and H\,{\small{II}}~regions.  The list of H\,{\small{II}}~regions is taken from the WISE catalog of H\,{\small{II}}~regions \citep{2014ApJS..212....1A}, and that of the SNRs from the latest version of Green's catalog \citep{2019JApA...40...36G}. Accurate spectral index determination based on our uGMRT data is hampered by the facts that the central frequencies of the two bands covered in our survey are not very different (400~MHz and 650~MHz), the data in the two bands have a non-trivial overlap of $uv$-coverage, and the flux density undetected due to the missing short-spacing information (see also \S\ref{subsec:missFD}).  Nevertheless, we see the spectral indices of SNRs to be significantly lower than those of H\,{\small{II}}~regions, which is what we expect to find.

The measured flux densities of SNRs detected in our images are presented in Table\,\ref{tab:FD} compared with expected values extrapolated from the 1.4~GHz flux densities from the THOR+VGPS data \citep{2016A&A...595A..32B} and the broad-band spectral indices derived by \citet{2023A&A...671A.145D}. Large-scale emission is filtered out by the uGMRT, while the THOR+VGPS data have zero-spacing information added. Consequently, the expected flux densities are anticipated to be larger than the observed values, which is indeed found to be the case for all the remnants except for G31.9+0.0.  In band-4, its measured and expected flux densities are equal within uncertainties, but in band-3, they are significantly different.  This is perhaps because the spectral indices were calculated by \citet{2023A&A...671A.145D} using a single power-law spectrum.  While G31.9+0.0 is known to show a turnover in its spectrum below 100~MHz \citep{2005AJ....130..148B}, there is a wide variation in the reported flux densities below 1~GHz as well \citep[see][]{1989ApJS...71..799K,1992AJ....103..943K,2005AJ....130..148B,2011A&A...536A..83S}.

\begin{table*}
\caption{Supernova remnant flux densities in the MeGaPluG pilot.}
\label{tab:FD}
\centering
\begin{tabular}{c r c c}
\hline\hline
name &  radius & $S_{\mathrm{400 MHz}}$ (Jy) & $S_{\mathrm{650 MHz}}$ (Jy) \\
~ & (arcmin) & Observed / Expected & Observed / Expected \\
\hline
G29.6$+$0.1 & 3.3 & 0.39$\pm$0.26 / 0.90$\pm$0.14 &  0.28$\pm$0.17 / 0.70$\pm$0.16 \\
G29.7$-$0.3 & 2.7 & 17.2$\pm$1.7 / 16.6$\pm$1.9  & 10.7$\pm$1.1  / 11.9$\pm$2.0 \\
G31.9$+$0.0 & 4.5 & 39.2$\pm$4.0 / 27.6$\pm$2.9  & 24.0$\pm$2.4  / 22.5$\pm$3.1 \\
G32.8$-$0.1 & 11.5& 10.1$\pm$1.3 / 15.3$\pm$1.7  &  3.6$\pm$0.5  / 13.4$\pm$1.8 \\
G33.2$-$0.6 & 9.2 & 1.7$\pm$0.4  / 4.2$\pm$0.5   &  1.1$\pm$0.2  / 3.6$\pm$0.5 \\
G33.6$+$0.1 & 6.7 & 19.3$\pm$2.1 / 20.4$\pm$2.3  &  12.9$\pm$1.3 / 16.0$\pm$2.4 \\
G34.7$-$0.4 & 19.2& 180$\pm$18   / 319$\pm$38    & 107$\pm$11    / 262$\pm$42 \\
\hline
\end{tabular}
\tablefoot{The expected flux densities in the two bands are calculated using the spectral indices derived in \citet{2023A&A...671A.145D}.}
\end{table*}

We searched for the recently identified SNR candidates in the THOR and GLOSTAR surveys \citep{2017A&A...605A..58A,2021A&A...651A..86D} in our images, but at a $3\sigma$-level, we cannot positively identify any of the 15 candidates that are located in the covered region.  Considering the fact that SNRs generally do not show spectral indices steeper than ${\sim}-0.8$, and that the largest angular scale detected in our data is less than $10'$, this inability to identify their counterparts in our images is not unexpected.  The sensitivity to extended emission must be generally better than 1--2 mJy beam$^{-1}$ in our data in order to detect these faint SNR candidates.  For the PWN-like GLOSTAR SNR candidate G031.256-0.041 and THOR SNR candidate G32.22-0.21, we are able to derive a lower limit of the spectral index (using the average surface brightness) of ${\sim}-0.1$ and ${\sim}-0.5$, respectively.  We also searched for shell-shaped objects that may have been undetected in the THOR and the GLOSTAR surveys because of synchrotron ageing at higher frequencies, but we do not find any such objects.  Longer integration time is required to constrain the spectral index for more SNR candidates and the identification of new candidates.

\section{Future work}
\label{sec:conclusions}

Recently, based on the new and the revised distances to several SNRs, \citet{2022ApJ...940...63R} provided evidence that the number of SNRs in the Milky Way must be at least over 2000 and may be as high as 5000, and not just 1000 as proposed by \citet{1991ApJ...378...93L}.  These numbers are glaringly inconsistent with that of the only  400 SNRs detected so far \citep{2019JApA...40...36G,2012AdSpR..49.1313F}.  Since a large fraction of the SNRs are expected to be only a few arcminutes in extent \citep{2023A&A...671A.145D}, and since SNRs are brighter at lower frequencies, significant progress can be made toward solving the problem of `missing' SNRs by a high-resolution, sub-GHz, and sensitive survey of the Milky Way.  Such a survey will also be useful to understand the spectral turnover caused by synchrotron self-absorption observed in many SNRs \citep[e.g.,][]{2011A&A...536A..83S,2023A&A...671A.145D}, as the turnover frequency is generally lower than 1~GHz.  In addition, one can also study the nonthermal emission that arise from star-forming regions \citep{2018MNRAS.474.3808V,2019MNRAS.482.4630V}.  The wide-band receivers on the uGMRT appear promising for a large-scale survey of the Milky Way. The promise of detecting SNRs in such low frequency surveys is illustrated by the fact that \citet{Brogan2006} detected 35 new SNRs in a 44 square degree area of the Galactic plane at 75 MHz (90 cm) with the classic VLA.

We performed this pilot study as a precursor to a larger survey of the Milky Way, the Metrewave Galactic Plane with the uGMRT (MeGaPluG) Survey, which we wish to undertake in the coming years.  We have developed automated calibration and imaging pipelines to reduce the large amount of data in an efficient manner.  The primary goal was to study the feasibility of imaging extended structures with the wide-band uGMRT at a high resolution, using only snapshot observations. Despite losing sensitivity and resolution due to the problem of non-working antennas, as is evident from Figure \ref{fig:b3b4mos}, we recover extended emission reasonably, which makes this a sensitive study for imaging extended structures at low frequencies ($<1$\,GHz) and high resolution ($25''$).

Based on the experiences from this study, for the upcoming larger survey, we design an optimum strategy for overcoming the shortcomings of the pilot study:
\begin{itemize}
    \item Undertake multiple scans of each field instead of just two scans as was done for this work.  The $uv$-coverage can be improved if we use four scans of 4--5 minutes spread over an observing session.  This also helps in minimizing the $uv$-coverage loss due to non-functional antennas during the observation.  
    \item The pointing configurations chosen (see Fig.~\ref{fig:pointings}) were not optimal given the size of primary beams. 
    A denser grid of pointings, as was done in the GLOSTAR survey \citep{2021A&A...651A..85B}, will result in better sensitivity although this will increase the observation time.
    \item In targeted observations, one can afford to observe the gain calibrator only once in an hour or so, since self-calibration is almost always guaranteed to enhance the dynamic range.  However, in our snapshot survey mode, self-calibration results in only a marginal improvement.  Hence, it may be wise to observe the gain calibrator more often, maybe once every half an hour.
    \item While polarization data are incredibly useful to study nonthermal emission, it is not possible to obtain any meaningful Stokes $Q$ and $U$ images of the target fields in our snapshot observations.  The polarization angles of sources with $|\mathrm{RM}|>7$~rad\,m$^{-2}$ rotate over the 200\,MHz bandwidth, causing depolarization.  While this limitation may be overcome by dividing the bandwidth into multiple frequency bins, the resulting deterioration of the signal to noise ratio makes the images resulting from the deconvolution process unusable.  Also, given that a large number of sources are located away from the center of the pointing, it is necessary to know the position dependence of the instrumental polarization effects, which has not yet been investigated for the uGMRT.  Hence, it is prudent to image and study only the Stokes~$I$ data under these circumstances.  However, since the cross-hand polarizations are helpful in eliminating the RFI, they may still be recorded.
\end{itemize}

\begin{acknowledgements}
    We thank the anonymous referee for their comments and suggestions.  RD is a member of the International Max Planck Research School (IMPRS) for Astronomy and Astrophysics at the Universities of Bonn and Cologne.  NR acknowledges support from Max-Planck-Gesellschaft through Max-Planck India partner group grant.  This research has made use of NASA's Astrophysics Data System and the SIMBAD database.  We have used the softwares Astropy \citep{2013A&A...558A..33A}, APLpy \citep{2012ascl.soft08017R}, Montage \citep{2003ASPC..295..343B}, and DS9 \citep{2003ASPC..295..489J} at various stages of this research. 

\end{acknowledgements}

\bibliographystyle{aa.bst}          
\bibliography{ref}

\clearpage
\begin{appendix} 

\section{Calibration scheme}
\label{apdx:calscheme}

\begin{enumerate}

\item{\it FITS to CASA Measurement Set (MS) conversion:}
The data for each observing session are provided on the NAPS website in a standard FITS format, which is converted to a MS using the CASA task {\tt importgmrt}.

\item{\it Hanning smoothing:}
In order to prevent Gibbs ringing, the task \texttt{hanningsmooth} is used to smooth the data with a Hanning window.

\item{\it Initial flagging:}
`Flagging' is the process of masking the data that cannot be recovered (generally due to poor instrument performance or RFI).  The task {\tt flagdata} is used for this purpose.  We flag dead antennas, shadowed antennas, and edge channels with low amplitude gains. 
Then, using the {\tt tfcrop} mode of the task {\tt flagdata}, we perform a round of automated RFI excision.  The {\tt tfcrop} mode uses an algorithm that identifies and flags outliers on the 2D time-frequency plane based on local statistics.  It can operate on un-calibrated data as it can account for the bandpass shape, and it is especially helpful in flagging short duration RFI and time-persistent narrow-band RFI.

\item{\it Set models:}
The task {\tt setjy} is used to set the model of the primary flux calibrator, 3C\,286, based on the flux density scale of \citet{2017ApJS..230....7P}.  The secondary calibrator, J1851+005, is initially assumed to be a point source with an amplitude 1\,Jy before a model of the field is produced.

\item{\it Delay and bandpass solutions:}
Phase-corrected data of the primary flux calibrator are used to derive the corrections due to instrumental delays and the bandpass shape.

\item{\it Apply, flag, and repeat twice:}
The delay and bandpass solutions are applied to the calibrator fields and the corrected data are flagged with {\tt tfcrop}.  New solutions are then derived keeping these flags.  This was done twice in order to ensure that the calibration tables are obtained only from reliable data.

\item{\it Prepare for gain calibration:}
The final calibration tables for delay and bandpass are applied on the primary and secondary gain calibrators.  Their corrected data are split to a new MS with appropriate channel-averaging.  This step significantly speeds up\footnote{All GMRT data are taken in spectral line mode (typically 8192 or 16384 channels) in order to isolate the narrow-band RFI which is ubiquitous below 1\,GHz at the observatory.  Because of the large number of channels, applying delay and bandpass tables `on the fly' (see \url{https://casadocs.readthedocs.io/en/stable/notebooks/synthesis\_calibration.html?highlight=fly\#Solve-for-Calibration}) in order to obtain temporal gain solutions takes a much longer time compared to the method we used.} the upcoming temporal gain calibration steps.

\item{\it Gain calibration:}
Using the data of primary and secondary calibrator fields, time-dependent solutions are derived for gain amplitude and phase variations caused by instrumental response and ionospheric effects.  The gain calibration tables are applied to the secondary calibrator field, and outliers in the corrected data column are flagged using the {\tt tfcrop} mode of the task {\tt flagdata}.  New gain solutions are obtained keeping the latest flags.  The amplitude corrections to the secondary calibrator are then scaled using the gains from the primary calibrator, with the task {\tt fluxscale}.

\item{\it A better gain calibration by modeling the secondary calibrator:}
If the secondary calibrator has been modeled as a point source, then a correct model needs to be made by self-calibration\footnote{The CLEAN parameters used for the self-calibration of the phase calibrator field are the same as those for target fields, and are discussed in the imaging section.}.  The CLEAN components of the final self-calibrated image are used as the model of J1851+005.  This step is done in order to account for the nearby extended sources in the model of the gain calibrator field.  New gain solutions are obtained using this model of the gain calibrator field.

\item{\it Apply calibration tables:}
The latest set of tables are then applied to the target fields.  The time-dependent gain solutions are linearly interpolated between adjacent scans, and the data that could not be calibrated and those with low signal-to-noise ratio are flagged.

\item{\it Flag RFI on targets:}
We perform a step of automated flagging on the target fields with the task {\tt flagdata} in the modes {\tt tfcrop} and {\tt rflag}.

\item{\it Split target fields:}
The calibrated data of target fields are split into separate MSs with the task {\tt split} and averaged in bandwidth in order to make the imaging process faster and convenient.  The data are averaged only to some extent such that no bandwidth smearing occurs (channel width must be ${\lesssim}$0.7\,MHz in the observed frequency regime).

\end{enumerate}

\section{Beam size estimation}
\label{apdx:beamsize}

The point spread function (PSF, dirty beam, or $B_\mathrm{dirty}$) of an interferometric observation depends on the array configuration and the sky position of the target being observed.  The raw image (dirty image, or $I_\mathrm{dirty}$) made by an interferometer before the CLEAN process is the sky brightness distribution ($I_\mathrm{sky}$) convolved with the PSF.  The final reconstructed image $I_\mathrm{reconstr}$ is made by adding the residuals ($I_\mathrm{res}$) to the convolution of a Gaussian CLEAN beam $B_\mathrm{clean}$ with the model of the sky image obtained by deconvolving the PSF from the dirty image, where the Gaussian CLEAN beam is estimated by fitting a Gaussian to the central lobe of the dirty beam.  In effect, the dirty beam is replaced by the CLEAN beam.  
\begin{equation}
\begin{aligned}
    I_\mathrm{dirty} &= I_\mathrm{sky} \circledast B_\mathrm{dirty} \\
                     &= I_\mathrm{model} \circledast B_\mathrm{dirty} + I_\mathrm{res} \\
    I_\mathrm{reconstr} &= I_\mathrm{model} \circledast B_\mathrm{clean} + I_\mathrm{res}
\end{aligned}
\end{equation}

In CASA, all the above is taken care of by a single task {\tt tclean}.  In the version 5.8, a major upgrade was made to the PSF fitting algorithm\footnote{see \url{https://casadocs.readthedocs.io/en/v6.2.0/notebooks/introduction.html}}, which changes the results of the Gaussian CLEAN beam fitting.  The images from the observations of May 27 during which one arm of antennas was completely flagged are drastically affected by this change.  The older CASA versions before 5.8 (ie $\le$5.7) gave the beam size for the May 27 observations as $\sim9''\times 8''$, whereas we obtain a size of $\sim 20''\times 5''$ when using v5.8.  The result from v5.7 is incorrect because the southern arm was completely flagged during that observation and the PSF must be highly asymmetric for snapshot observations such as ours. 

To make sure that this discrepancy was not a result of improper calibration, we imaged a simulated GMRT dataset made with CASA simulations without any corrupting gains.  We flagged the antennas on one arm and imaged the data with the same parameters in AIPS, CASA v5.7 and CASA v5.8.  The results were similar to the beam sizes obtained in the real data, and hence we confirm that the PSF fitting algorithm used in AIPS and older versions of CASA delivers incorrect results when applied to our observations.  All imaging took place in CASA v5.8 in order to ensure that the beam sizes are correctly estimated.  Incorrect beam size estimation leads to unnatural flux densities of imperfectly deconvolved sources.

\end{appendix}

\end{document}